# A Resilient and Energy-Efficient Smart Metering Infrastructure Utilizing a Self-Organizing UAV Swarm

Mustafa Siham     Qutaiba I. Ali
University of Mosul
Iraq

**Abstract:** The smart metering infrastructure may become one of the key elements in efficiently managing energy in smart cities. At the same time, traditional measurement record collection is performed by manual methods, which raises cost, safety, and accuracy issues. This paper proposes an innovative SMI architecture based on an unmanned aerial vehicle swarm organizing itself for the autonomous data collection in smart metering infrastructure with scalability and cost-effectiveness while minimizing risks. We design an architecture-based comprehensive system with various phases of operation, communication protocols, and robust failure-handling mechanisms to ensure reliable operations. We further perform extensive simulations in maintenance of precise formations during flight, efficient data collection from smart meters, and adaptation to various failure scenarios. Importantly, we analyze the energy consumption of the proposed system in both drone flight operations and network communication. We now propose a battery sizing strategy and provide an estimate of the operational lifetime of the swarm, underlining the feasibility and practicality of our approach. Our results show that UAV swarms have great potential to revolutionize smart metering and to bring a further brick to greener and more resilient smart cities.
**Keywords:** Smart City, UAV Swarm, Smart Metering, Self-Organizing Network, Failure Recovery, System Design, Energy Consumption

## 1. Introduction

The rapid urbanization worldwide demands a smart, innovative, and sustainable application of resources, mitigation within environmental impacts, and enhancements in citizen livelihood. In this regard, connected with integrated infrastructure and with data-driven decision processes, the smart city has emerged as an important development paradigm for cities. While this is a very important basis for a smart city, SMI facilitates smart metering: it effectively provides utilities and consumers with real-time and accurate energy consumption data that empowers these two stakeholders to work together toward better energy use patterns for a more sustainable energy future [1]. While SMI might have its potential for transformation, traditional architectures have been highly relying on manual data collection methods, with human personnel actually going from door to door or building to building for the collection of data from the nescars. There are a number of inherent limitations to this reliance on manual processes:
- High Operational Costs: The employment of field personnel in collecting data means higher salary, transportation, and administrative overhead costs. These costs rise even more during the deployment of personnel in difficult geographical terrains or areas that are highly inhabited by people.
- Safety Risks to Personnel: Deployment of human workers in hazardous environments, such as unstable infrastructures, extreme weather conditions, and insecurity are a huge safety risk to personnel. Manual data collection exposes personnel to accidents, injuries, and health hazards.
- Data inaccuracies and delays: Most data intake mechanisms worked manually are invariably prone to human inaccuracies, further bringing down the accuracy of the consumption readings. Besides that, the time it takes for the actual collection of data manually from a large number of locations inherently makes data availability delayed and limits the real value of real-time monitoring and analysis. Recent developments in Unmanned Aerial Vehicle, or drone, technology have opened exciting vistas in use for transforming many industries and sectors, including urban infrastructure management. Indeed, UAVs inherently offer a very attractive alternative to manual data collection in SMI for several reasons:
- Agility and Maneuverability: It can navigate through complicated urban environments to reach places where it is hard or dangerous for human personnel. Their agility to move through narrow spaces and to fly at different altitudes favors them as an optimal choice for data acquisition in a wide range of settings.
- Scalability and Cost-Effectiveness: The deployment of a swarm of interconnected UAVs will have the added advantage of rapid data collection in a very short time from a large number of smart meters deployed within a wide area. Such a scalable solution reduces the time and cost required for manual data gathering considerably.
- Reduced Safety Risks: The use of UAVs in data collection will reduce the need to deploy personnel in hazard-prone environments, and by doing so it reduces accidents, injuries, or health hazards to workers significantly.
- Improved Data Accuracy and Timeliness: Equipped with appropriate sensors, UAVs would be able to download data directly from smart meters with no potential human error from manual entry. Besides, rapid deployment and data transmission by UAVs provide real-time monitoring and analysis in a timely manner.

Given the potential of overcoming limitations of conventional SMI architectures with a swarm of UAVs, in this paper we present a novel, resilient, and autonomous SMI system that leverages the power of collaborative UAVs. Our main research objectives in this respect are:

1. The design and development of a complete SMI architecture based on a self-organizing swarm of UAVs for smart meter data collection in an efficient and reliable manner. This architecture will involve designing DMC, UAV swarm, communication protocol, and operational phases that will work seamlessly in data acquisition.

2. Embed robust failure handling mechanisms that ensure continuity of operations or mission success in case of malfunction by either of the drones, unexpected events occurring, or specifically under faulty environmental conditions. It describes how to deal through estimated and sudden failures of drones, mechanisms for dynamic task reassignment to ensure continuity in data collection.

3. Through extensive simulations that emulate various different real-world use cases, this paper establishes the efficacy of the proposed system. These shall check formation accuracy, efficiency in data gathering, recovery times given certain failure events, and system-wide resilience in duress.

The contribution of this research work is bound to add to the increasing literature in current knowledge on UAV applications towards smart city development. Our proposed SMI architecture above presents a practical and scalable solution for improving efficiency, reliability, and safety of smart metering that shall lead to more sustainable and resilient urban environments.

**2. Related Work**

Many research works have been published to facilitate the advancement of UAV in various fields as follows:
2.1 Applications of UAVs in Smart Cities
In recent years, several studies have explored using UAVs for different smart city applications. These applications are wide-ranging, including [4-7]:
1. Traffic Monitoring and Management: UAVs equipped with cameras and sensors can provide real-time traffic flow data, allowing public service agencies to respond faster than usual.
2. Environmental Monitoring and Pollution Control: UAVs can monitor air quality, detect pollution sources, and assess urban vegetation health, helping to build healthier living environments.
3. Infrastructure Inspection and Maintenance: UAVs with high-resolution cameras and thermal imaging are employed to inspect bridges, power lines, and other vital infrastructure segments for damage, preventing failures.
4. Public Safety and Security: UAVs can assist in surveillance, search and rescue, and crowd control, enhancing public safety.

2.2 UAV Swarm Technology
While deploying single UAVs has benefits, coordinating multiple UAVs as a collaborative swarm can unlock even more potential for tasks requiring scalability, adaptability, and resilience. Based on the collective behavior seen in nature (e.g., bird flocks, fish schools, insect swarms), UAV swarm technology involves managing interactions among autonomous vehicles. Key features of UAV swarms include [8-12]:
1. Decentralized Control: Swarms don't rely on a central controller, instead using local interactions between individual UAVs to cooperate and achieve group goals, enhancing resilience.
2. Self-Organization: UAV swarms adapt to changing conditions and reconfigure themselves autonomously, essential for dynamic environments.
3. Scalability and Fault Tolerance: Swarm size can adjust based on mission needs, and the system is fault-tolerant. If one UAV fails, others compensate to continue the mission.

2.3 UAV-Based Smart Metering Systems
Research has also explored UAVs for smart meter data collection, with different approaches [13-17]:
1. Single UAV Deployment: A single UAV visits each smart meter to collect data. Although simple, this method lacks scalability for large-scale deployments.
2. Static UAV Deployment: UAVs are positioned at fixed points to collect data, eliminating the need for continuous flight but may not suit dynamic data needs.
Our proposed system leverages a self-organizing UAV swarm to overcome these limitations, providing scalable, efficient, and resilient smart meter data collection.

## 3. System Design and Architecture

### 3.1 System Overview
The proposed UAV swarm-based SMI system comprises two primary components, see Figure 1:
1. Data Management Center (DMC): Acts as the central command and control hub for the entire system. It is responsible for mission planning, swarm configuration, data processing and analysis, and communication with external systems.
2. UAV Swarm: Consists of a designated Leader Drone (LD) and multiple Slave Drones (SDs). The LD coordinates the swarm's actions, relays instructions from the DMC to the SDs, aggregates data collected by the SDs, and transmits it back to the DMC. The SDs are responsible for collecting data from individual smart meters using their onboard sensors.

### 3.2 Data Management Center (DMC)
The DMC serves as the central nervous system of the SMI architecture, responsible for:
1. Mission Planning and Configuration: Human operators interact with the DMC to define mission parameters, such as the target area for data collection, desired swarm formation, data collection frequency, and any specific waypoints or flight paths the swarm should follow.
2. Swarm Configuration and Management: The operator selects the number of SDs to deploy, assigns a specific drone as the LD, and monitors the status of individual drones within the swarm through a user-friendly interface.
3. Data Processing and Analysis: The DMC receives data collected by the UAV swarm, processes it to extract relevant information, performs analysis to identify consumption patterns or anomalies, and generates reports for further action or decision-making.
4. Communication and Interfacing: The DMC establishes communication links with the LD using a reliable wireless communication protocol. It also interfaces with external systems, such as utility company databases or energy management platforms, to share data and facilitate integration with existing infrastructure.

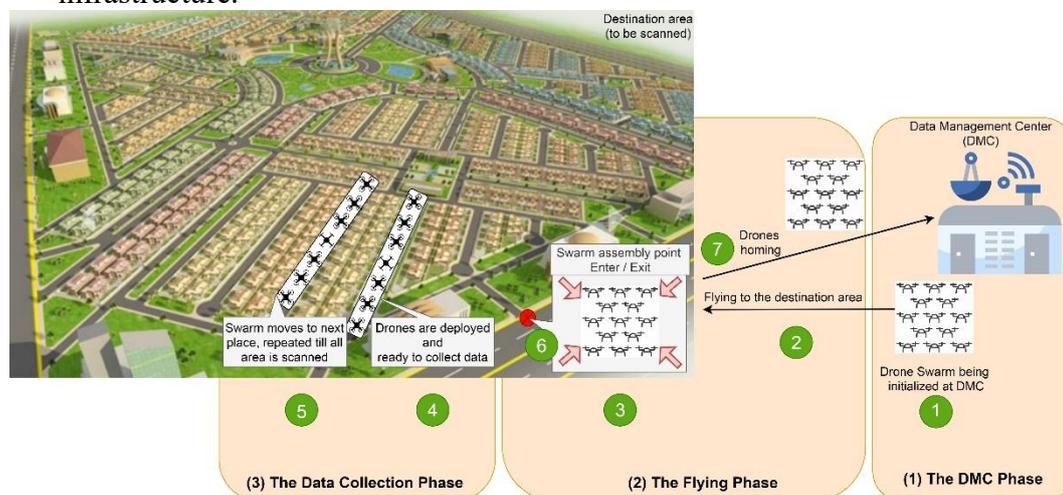

Figure 1: The proposed system procedural diagram.

### 3.3 UAV Swarm: Leader Drone (LD) and Slave Drones (SDs)

### 3.3.1 Leader Drone (LD)

The LD plays a critical role in swarm coordination and communication. It is responsible for:

1. Receiving Mission Instructions: The LD receives detailed mission parameters from the DMC, including GPS waypoints, desired swarm formation, and data collection instructions.
2. Relaying Instructions to SDs: The LD disseminates mission instructions received from the DMC to the SDs, ensuring synchronized swarm movements and task execution.
3. Monitoring Swarm Status: The LD continuously monitors the status of individual SDs, including battery levels, location, and sensor readings. It relays this information back to the DMC for real-time situational awareness.
4. Aggregating Data from SDs: As SDs collect data from smart meters, they transmit it to the LD, which aggregates the data from all SDs and periodically transmits it back to the DMC for processing and analysis.

### 3.3.2 Slave Drones (SDs)

The SDs are the workhorses of the system, responsible for:

1. Following LD Instructions: SDs receive and execute instructions from the LD, maintaining formation during flight, deploying to specific locations, and initiating data collection sequences.
2. Data Collection from Smart Meters: SDs are equipped with appropriate sensors to collect data from smart meters. This may involve Optical Character Recognition (OCR) to read data from traditional meters or wireless communication protocols to interface with smart meters directly.
3. Transmitting Data to the LD: Once data is collected, SDs transmit it wirelessly to the LD for aggregation and eventual transmission to the DMC.

### 3.4 Communication Protocols and Data Exchange

Efficient and reliable communication between the DMC, the LD, and the SDs is crucial for successful swarm operation.

- DMC to LD Communication: We propose utilizing a robust wireless communication protocol, such as 4G/LTE or potentially 5G in areas with coverage, for communication between the DMC and the LD. This ensures a stable connection with sufficient bandwidth for transmitting mission data and receiving status updates and aggregated data from the swarm.
- LD to SD Communication: For communication between the LD and SDs, a suitable wireless local area network (WLAN) protocol, such as Wi-Fi or Zigbee, can be employed. These protocols offer high data rates, low latency, and energy efficiency, essential for maintaining swarm coordination and exchanging data effectively.

### 3.5 Operational Phases

The operational cycle of the proposed UAV swarm-based SMI system can be broken down into three distinct phases, see Figure 2:

### 3.5.1 DMC Phase (Initialization and Configuration):

1. Swarm Power-Up: The operator powers up the required number of SDs and the designated LD.

2. Connection Establishment: The LD establishes a secure connection with the DMC, and the SDs connect to the LD, forming the swarm network.
3. Mission Information Upload: The operator defines mission parameters (e.g., target GPS coordinates, swarm formation, data collection frequency) through the DMC interface. The DMC transmits this information to the LD.
4. SD Configuration: The LD receives the mission information and configures each SD with its specific role and tasks for the mission.

3.5.2 Flying Phase (Transit and Formation):
1. Swarm Launch: Upon receiving the launch command from the DMC, the LD initiates takeoff, followed by the SDs.
2. Formation Establishment: The LD guides the SDs to form the pre-defined swarm formation (e.g., linear, grid) while navigating towards the target area.
3. Waypoint Navigation: The LD, following the designated flight path and waypoints, leads the swarm to the target location for data collection. The SDs maintain their relative positions within the formation throughout the flight.

3.5.3 Data Collection Phase (Deployment and Sensing):
1. Target Area Arrival: The LD, upon reaching the designated target area, signals the SDs to prepare for deployment.
2. SD Deployment: The SDs autonomously deploy to their assigned locations within the target area, following a pre-defined deployment strategy to ensure efficient coverage of smart meters.
3. Data Acquisition: SDs activate their onboard sensors and collect data from the designated smart meters. The data collected may include energy consumption readings, voltage levels, and other relevant parameters.
4. Data Transmission to LD: Each SD transmits its collected data wirelessly to the LD for aggregation.
5. LD Aggregation and Transmission to DMC: The LD aggregates the data received from all SDs and periodically transmits it back to the DMC using the established long-range communication link.
6. Mission Completion and Return: Upon receiving confirmation from the DMC that sufficient data has been collected, the LD initiates the swarm's return to the launch site, maintaining formation throughout the flight back.

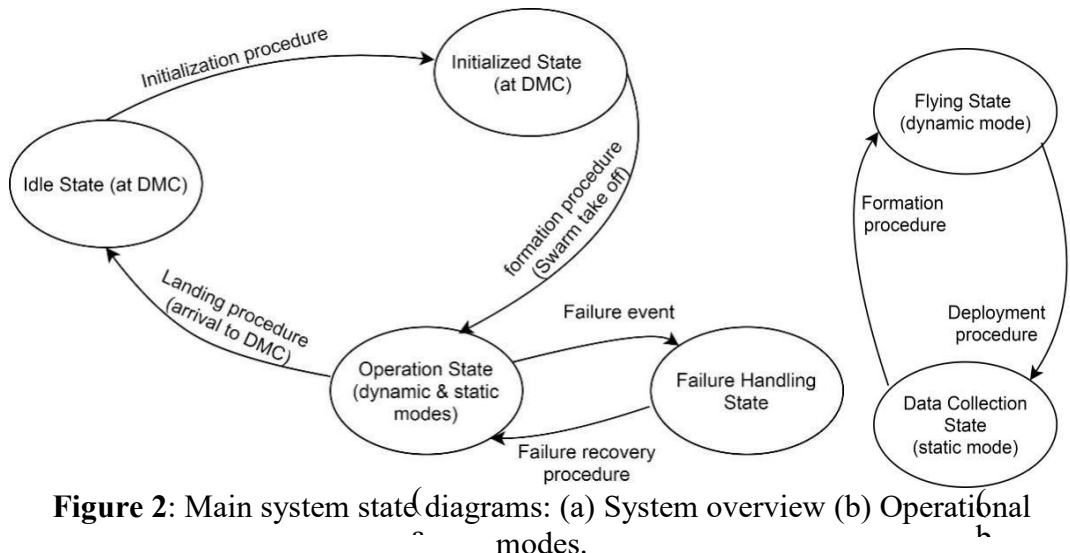

**Figure 2**: Main system state diagrams: (a) System overview (b) Operational modes.

## 4. Ensuring System Reliability: Robust Failure Handling

Real-world deployments of UAV swarms are susceptible to various uncertainties, including drone malfunctions, communication interruptions, and unpredictable environmental conditions. To ensure system reliability and mission success even in the presence of such challenges, we have incorporated robust failure-handling mechanisms into our design, see Figure 3:

4.1 Leader Drone (LD) Failure Handling

LD failure poses a significant risk to mission success as it acts as the central coordinator of the swarm. To address this, we implement a multi-layered approach:
1. Backup LD Designation: During the initialization phase, the operator designates a specific SD as the backup LD. This backup LD possesses all the capabilities of the primary LD and remains on standby throughout the mission.
2. Hard Handover (Immediate LD Failure): In the event of a sudden and unexpected LD failure (e.g., collision, loss of communication), the backup LD immediately takes over the leadership role. It assumes responsibility for swarm coordination, data aggregation, and communication with the DMC, ensuring minimal disruption to the mission.
3. Soft Handover (Predicted LD Failure): To further enhance resilience, we introduce a novel failure prediction mechanism. The LD continuously monitors its onboard sensors (e.g., battery level, temperature) and can predict potential failures in advance. If the LD anticipates a failure, it initiates a soft handover process, transferring leadership to the designated backup LD before the failure occurs. This proactive approach allows for a smoother transition and potentially extends the operational life of the failing LD by allowing it to enter a power-saving mode.

4.2 Slave Drone (SD) Failure Handling

While SD failures are less critical compared to LD failure, they can still impact overall mission efficiency. To address SD malfunctions, we implement:
1. Dynamic Task Reallocation: If an SD fails or becomes unresponsive, the LD dynamically reassigns its tasks to other available SDs within the swarm. This ensures that all designated smart meters are covered, maintaining data collection continuity.

2. Failure Isolation: The LD isolates the failed SD from the swarm network to prevent potential communication interference or disruption to other operational drones.
3. Optional Return to Base: Depending on the severity of the failure and mission parameters, the LD can instruct the failed SD to return to the base station autonomously for maintenance or replacement.

4.3 Environmental Disturbance Mitigation

Operating in real-world urban environments exposes the UAV swarm to various environmental disturbances, such as wind gusts and obstacles. To mitigate the impact of such disturbances:

- Robust Formation Control Algorithm: The swarm utilizes a robust formation control algorithm that considers environmental factors and adjusts drone positions and movements dynamically to maintain formation integrity and prevent collisions.
- Onboard Sensors for Obstacle Avoidance: Each drone is equipped with onboard sensors (e.g., cameras, ultrasonic sensors) to detect and avoid obstacles autonomously, ensuring safe navigation through complex urban environments.

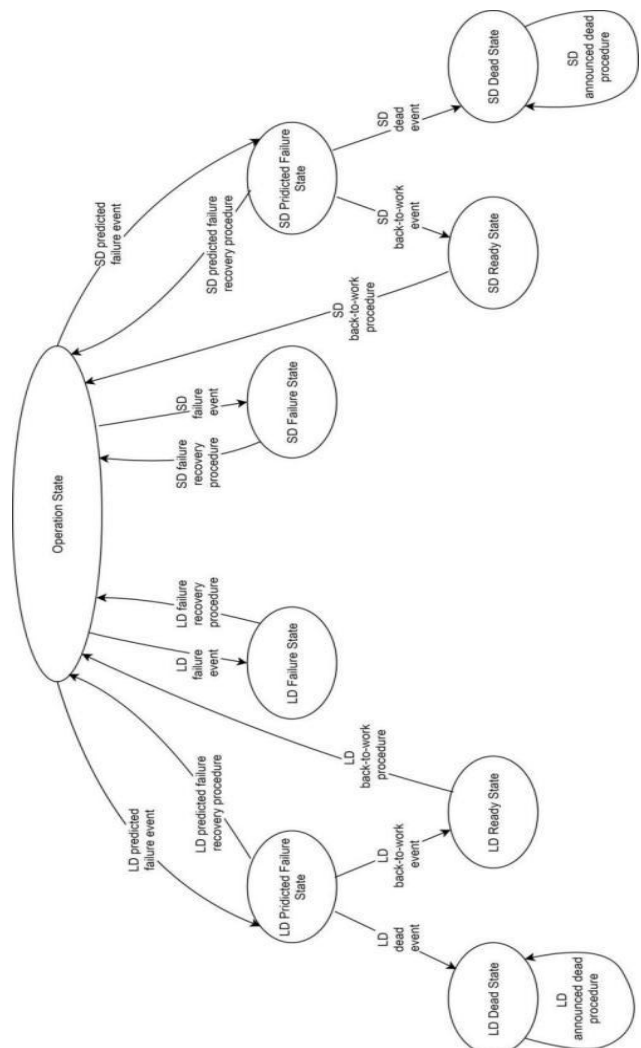

**Figure 3**: Failure Handling Diagram.

5. Performance Analysis and Implementation Issues

This section focuses on evaluating and validating the proposed Smart Metering Infrastructure (SMI) framework introduced in the previous chapters. The validation involves a real-world use case – a COVID-19 testing application – simulated using the OPNET network simulation tool. The section further explores energy consumption calculations and proposes a security model for the system.

5.1 The Proposed System in a Healthcare Application: A COVID-19 Use Case
We introduce a timely and relevant application of the proposed SMI framework: a Portable Health Clinic (PHC) system using a UAV swarm for automated COVID-19 testing during lockdown situations. This application highlights the system's potential to provide essential services with minimal human intervention and risk, especially in challenging scenarios, see Figure 4.

5.1.2 The Proposed Swarm-Based System Architecture
Figure 5.1 depicts the system architecture, featuring three operational layers:
- Drone Level: Consists of Slave Drones (SDs) collecting COVID-19 test data and a Leader Drone (LD) aggregating the information and communicating with the local clinic.
- Local Clinic Level (DMC): Manages multiple drone swarms, oversees data processing, and escalates critical cases to the general hospital.
- General Hospital Level: Provides the highest level of medical expertise, resources, and supervision for COVID-19 containment.

Figure 4 further illustrates the hierarchical distribution of roles within this architecture, emphasizing scalability and efficient management.

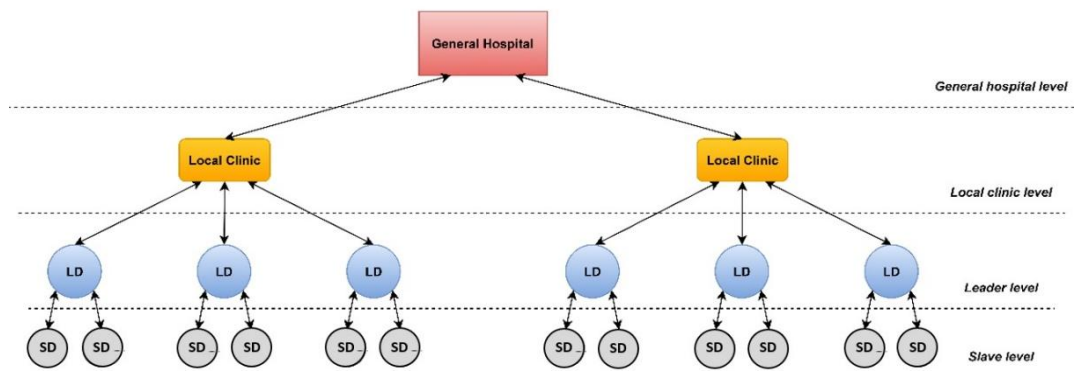

Figure 4: The suggested system layered architecture

5.2.3 The Proposed Swarm-Based System Description
Figure 5 illustrates the operational procedure of the PHC system. The system leverages a UAV swarm equipped with testing sensors to perform contactless COVID-19 screening at people's doorsteps. A triage system categorizes the case severity, and appropriate actions are taken, including alerting the local clinic, providing precautionary instructions, or initiating a video call with a doctor. This automated process enhances efficiency, reduces human risk, and enables timely medical intervention.

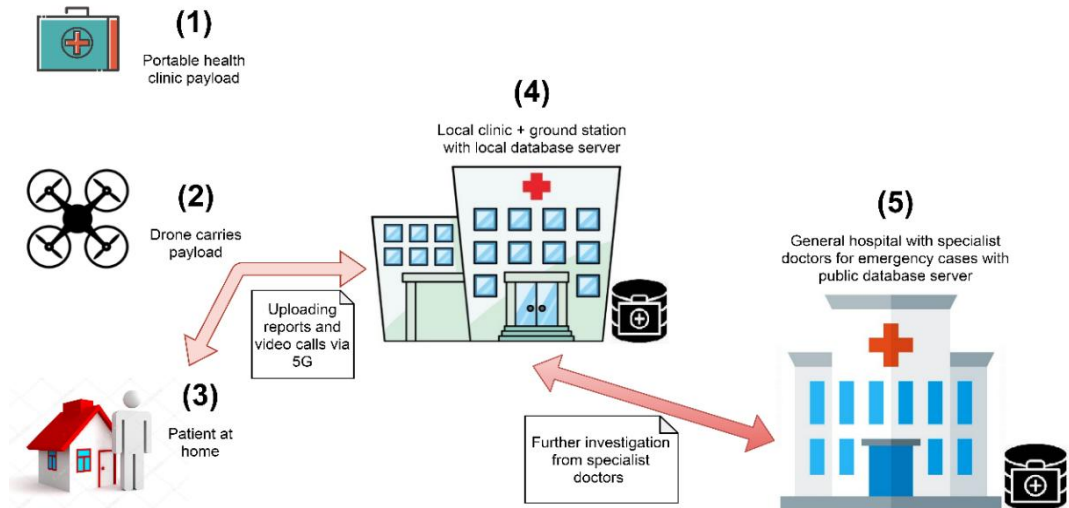

**Figure 5**: The proposed Portable Health Clinic (PHC) system operational procedure.

5.2.4 The Roles and Responsibilities of the Proposed Swarm-Based System Units
This section details the roles and responsibilities of each system unit, emphasizing hierarchical distribution and collaborative functionality:
- Local Clinic (DMC): Houses the drone swarm, manages mission information, receives swarm status reports, interacts with patients, and escalates cases to the general hospital when needed.
- Slave Drones (SDs): Perform individual testing procedures, collect vital signs data, apply the COVID-19 triage process, provide instructions to residents, and report cases to the LD.
- Leader Drone (LD): Coordinates the swarm, manages SD deployments, communicates with the DMC, and handles failure scenarios.
- Person Being Tested: Follows drone instructions, provides vital signs data, and participates in video calls with doctors if needed.
- General Hospital: Acts as the highest medical authority, provides specialist consultation, manages critical cases, and conducts contact tracing.

The system operates in two modes: dynamic (flying) and static (fixed), each with specific tasks and communication patterns.

5.2 The Simulation Model Using OPNET
The chapter presents a simulation model developed using the OPNET network simulation tool to validate the system's performance [18-22]. The model, using a map of Ashti City 1 in Erbil, Iraq, aims to realistically simulate swarm operation, communication patterns, and network behavior.

5.2.1 Assumptions and Initial Settings
The simulation model is built with several assumptions to simplify complexity while maintaining realism. Table 1 summarizes the initial settings of the simulation model, including simulation time, number of drones, network span area, distances between system components, and adopted technology parameters.

Table 1: The initial settings of the simulation model

| Simulation Parameter | Value |
|---|---|
| Simulation Time | 15 minutes |
| Number of drones | 1 to 100 (no video call) <br> 1 to 14 (video call) |
| Network span area | 2 Km x 2 Km |
| Distance between drones and formation | 12 meters with the linear formation |
| Distance between swarm and WiMax BS | 1 Km |
| Distance between swarm DMC and WiMax BS | 1 Km |
| 4G adopted technique settings | WiMax technology <br> Modulation and coding: 64-QAM 3/4 <br> Scheduling type: rtPS <br> Max. sustained traffic rate: 10 Mbps <br> Min. reserved traffic rate: 5 Mbps |
| WLAN adopted technique settings | 802.11a (OFDM) <br> Data rate: (6,18, 36, 54) Mbps <br> Node buffer size = 1M bit <br> packet processing rate=(5000,10000,20000) pkt/s <br> Block ACK: EDCA (802.11e) disabled/enabled <br> WLAN MTU = WiMAX MTU = 1500 byte |
| Swarm status | Landed, power-saving enabled and gathering data. |

5.2.2 Traffic Profiles

Table 2 describes the two main traffic profiles used in the simulation:
- Traffic Profile 1: Periodic status reports exchanged between SDs, the LD, and the DMC.
- Traffic Profile 2: Video call traffic, added when medical consultation is needed.

These profiles are designed to mimic realistic communication patterns within the system.

Table 2: Traffic profiles of the simulated drone swarm.

| Traffic Profile | Application | Description | |
|---|---|---|---|
| 1 | SD reporting status | (SD → LD: statusReport$_{SD}$) <br> Packet length = 13 byte <br> Packet rate = 0.1 packet/s | (LD→SD: ACK) <br> Packet length = 2 byte <br> Packet rate = 0.1 packet/s |
| | LD reporting status | (LD → DMC: statusReport$_{LD}$) <br> Packet length=12x(**nx12)byte <br> Packet rate = 0.033 | (DMC→LD: ACK) <br> Packet length = 2 byte <br> Packet rate = 0.033 packet/s |

| | | packet/s | |
|---|---|---|---|
| 2 | SD reporting status | (SD → LD: statusReport$_{SD}$) Packet length = 13 byte Packet rate = 0.1 packet/s | (LD→SD: ACK) Packet length = 2 byte Packet rate = 0.1 packet/s |
| | LD reporting status | (LD → DMC: statusReport$_{LD}$) Packet length = 12*(n*12) byte Packet rate = 0.033 packet/s | (DMC→LD: ACK) Packet length = 2 byte Packet rate = 0.033 packet/s |
| | Case reporting | (SD → LD → DMC: caseReport) Packet length = 500 byte Packet rate = event-driven | (DMC → LD → SD: ACK) Packet length = 2 byte Packet rate = event-driven |
| | Video call | (SD ←→ LD ←→ DMC: video conference) available resolutions: Bandwidth requirements = 2Mbit/s, 4Mbit/s, 6Mbit/s. Frame rate = 30 frame/s. | |

**n: Number of SDs in the swarm.

5.3 Results and Discussion
This section presents the simulation results and analyzes the system's performance based on several networking metrics. Here's an overview of the simulation scenarios and key results:

**Scenario 1: No Video Call**
- Focus: This scenario assesses the system's performance under normal operating conditions, where drones exchange periodic status reports (traffic profile 1) without any video calls.
- Variables: The simulation tests various data rates (54, 36, 18, and 6 Mbps), swarm sizes (1 to 100 UAVs), and packet processing rates (5000, 10000, and 20000 packets/second).
- Key Findings:
  - High Throughput, Primarily over WLAN: The system achieved high throughput, mainly attributed to the efficient exchange of status reports within the swarm over WLAN. WiMAX throughput was significantly lower as it only involved periodic communication between the LD and DMC.
  - Low Latency: The latency was consistently low (250-350 microseconds) due to minimal network traffic.
  - Scalability: Increasing the number of UAVs didn't significantly affect throughput or latency, indicating good system scalability when handling status reports.

**Scenario 2: Video Call Scenarios (2Mbit/s, 4Mbit/s, 6Mbit/s)**
- Focus: These scenarios evaluate the system's performance when handling video calls for remote medical consultations, adding significant data traffic to the network (traffic profile 2).

- Variables: Each video call scenario (2Mbit/s, 4Mbit/s, 6Mbit/s) is tested with varying data rates, a limited number of simultaneous video calls (14, 6, and 4 respectively), and different packet processing rates.
- Key Findings:
    - Throughput Bottleneck: WiMAX becomes the primary bottleneck, limiting the number of simultaneous video calls due to its lower bandwidth compared to WLAN.
    - Latency Impact: Latency increases significantly with higher video resolutions and lower data rates.
    - WLAN Efficiency: WLAN consistently demonstrated lower latency than WiMAX, especially at lower data rates.

**Scenario 3: Enabling EDCA**
- Focus: This scenario investigates the impact of enabling Enhanced Distributed Channel Access (EDCA), a QoS mechanism in 802.11e, on latency.
- Variables: The simulation compares latency with and without EDCA for different video resolutions at a fixed data rate (54 Mbps) and packet processing rate (10000 packets/second).
- Key Findings: EDCA effectively reduced latency by about 3%, particularly with higher traffic loads (more video calls). This improvement highlights the benefits of QoS mechanisms for prioritizing critical data traffic.

Additional Findings:
- Packet Data: The number of packets sent and received by SDs was almost identical due to the dominance of bidirectional video call traffic.
- Data Traffic: The total data traffic transmitted over WLAN was noticeably higher than WiMAX due to the additional headers required for WLAN frames.
- Packet Loss: The measured packet loss was less than 5%, considered acceptable for this application.

**6. Energy Consumption Analysis**

A critical aspect of designing a practical and sustainable UAV swarm-based SMI system is analyzing and optimizing energy consumption. In this section, we examine the energy expenditure associated with various phases of swarm operation and propose a battery sizing strategy to ensure sufficient mission duration.

6.1 Energy Consumption during Swarm Flight

We assume a distance of 1 km between the swarm's DMC and the target data collection area. With a flying speed of 12 km/hr in matrix formation, the swarm requires approximately 5-6 minutes to reach its destination, considering formation and deployment time.

6.1.1 Energy Consumed by Drone's Rotors

The selected drone model, DJI Phantom 4 Pro V2.0, has a battery capacity of 5870 mAh at 15.2V, providing approximately 89.2 Wh of energy. This translates to a maximum flight time of around 30 minutes without any payload. However, our system incorporates a payload of approximately 200g per drone, consisting of sensors, a processing unit, and additional batteries.

Based on research [16] on drone payload impact on power consumption and flight time, Figure 6 illustrates the relationship between payload weight, power consumption, and flight time.

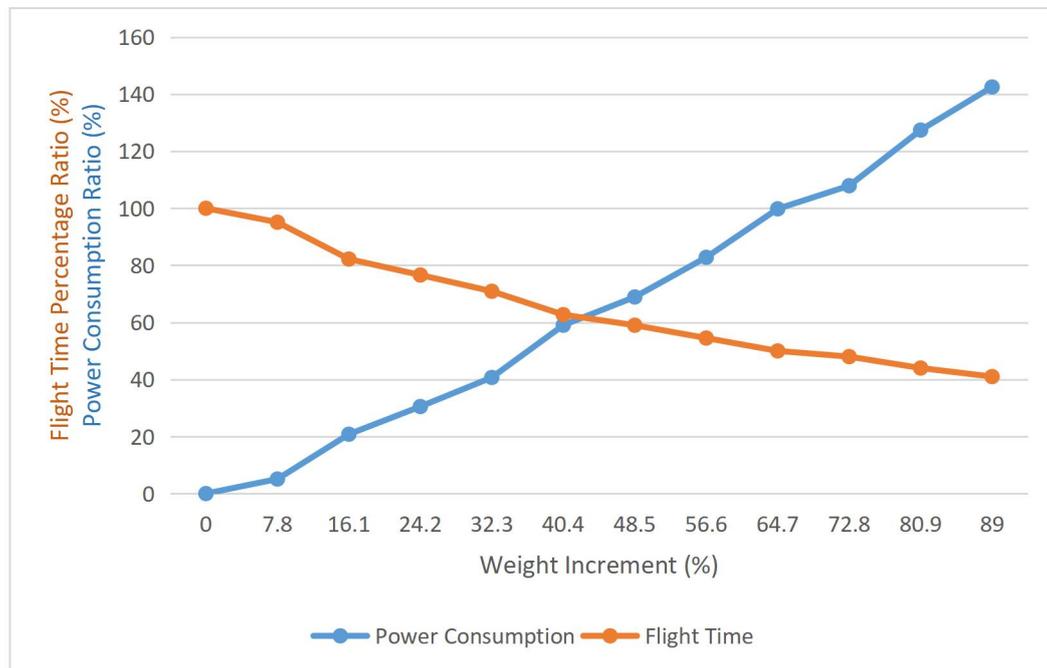

Figure 6: Drone's Payload Relation with Power Consumption and Flight Time

The additional payload weight increases power consumption and reduces flight time. Using equation (1), we calculate the payload-to-weight ratio:

Payload (%) = (Payload weight / Original drone weight) × 100% …………….. (1)

Table 3 presents the payload-to-weight ratios for both LD and SD.

Table 3: Payload-to-Weight Ratio

| Drone Type | Drone Weight (g) | Payload (g) | Payload-to-Weight Ratio (%) |
|---|---|---|---|
| LD | 1375 | 201.6 | 14.7 |
| SD | 1375 | 198 | 14.4 |

With a payload of approximately 15% of the drone's weight, Figure 6 indicates an increase in power consumption by about 20%, resulting in a 20% reduction in flight time. Consequently, the drones with payload have an estimated flight time of 24 minutes.
The total flight time for a mission, considering multiple data collection sessions and repositioning between target locations can be calculated using equation (2):

Total Flight Time = 2 × DMC Flight Time + (n × Repositioning Flight Time) ... (2)

Where 'n' represents the number of data collection sessions. Repositioning flight time is estimated to be 1 minute per session.

### 6.1.2 Power Consumed by Networking

During the flight, the swarm utilizes communication profile 1, as described in Section 5.3, involving status report messages between SDs, the LD, and the DMC. Additionally, the LD broadcasts the "MoveToWaypoint" message every 0.2 seconds to coordinate SDs during flight, with corresponding acknowledgments (ACKs).

Using OPNET simulations for a swarm with 1 LD and 10 SDs, we quantified the data traffic transferred during flight, as shown in Table 4.

Table 4: System Data Traffic (bps) during Flight

| From | To | Data Rate (bps) |
|---|---|---|
| SD | LD | 8549 |
| LD | SD | 9621 |
| LD | DMC | 44 |
| All SDs | LD | 85490 |
| DMC | LD | 10 |

### 6.2 Power Consumption during Swarm Data Collection

Upon reaching the target area and deploying, the SDs initiate data collection by recording video for 1 minute per person to assess vital signs [163, 164]. This data is processed onboard using MATLAB on a Raspberry Pi for case classification (healthy, suspicious, infected, emergency). In case of an infected or emergency case, a "caseReport" message is sent to the LD and DMC, potentially triggering a 5-minute video call for remote diagnosis.

We assume a 2%-3% infection rate in hotspot areas [15], leading to one case requiring escalation per swarm every 30 minutes. Each SD is assigned to a single house, with an assumed data collection time of 30 minutes per house.

### 6.2.1 Power Consumed When No Video Call

During normal operation with no escalated cases, only status reports are exchanged (profile 1). Table 5 presents the data traffic in this scenario.

Table 5: System Data Traffic (bps) during Data Collection with No Video Call

| From | To | Data Rate (bps) |
|---|---|---|
| SD | LD | 372 |
| LD | SD | 64 |
| LD | DMC | 44 |
| All SDs | LD | 3714 |
| DMC | LD | 10 |

### 6.2.2 Power Consumed with One Video Call

In this scenario, one SD classifies a case as infected or emergency, triggering a "caseReport" message and a 5-minute video call through the LD. Table 6 presents the data traffic for this session.

Table 6: Data Traffic (bps) during One Video Call

| From | To | Data Rate (bps) |
|---|---|---|
| SD | LD | 1144896 |
| LD | SD | 1144896 |
| LD | DMC | 1024533 |
| DMC | LD | 1024533 |

This video call requires a 2 Mbps bandwidth for both WiMAX and WLAN, leading to a significant increase in energy consumption.

6.3 Battery Sizing and System Lifetime

To determine the required battery capacity and system lifetime, we consider the energy consumption during flight, data collection, and repositioning, ensuring equation (3) is satisfied:

Battery Capacity > (2 × DMC Round Trip Energy) + ($N_{UAV}$ × (One Session Energy + Repositioning Flight Energy)) …………………………………………………………. (3)
Where '$N_{UAV}$' represents the number of data collection sessions per drone

Based on energy consumption calculations and considering a 30-minute session with one potential video call, Table 7 presents the estimated operational duration for different battery configurations.

Table 7: Swarm Operation Durability in Ideal Circumstances

| Battery Type | Max. Number of Sessions | Max. Operating Time (hours) |
|---|---|---|
| Drone Battery (LD) | 12 | 6 |
| Drone Battery (SD) | 12 | 6 |
| Raspberry Pi Battery (LD) | 28 | 14 |
| Raspberry Pi Battery (SD) | 15 | 7.5 |

The maximum operating time for the proposed system, limited by the drone battery, is 6 hours, encompassing 12 data collection sessions. Extending mission duration would require additional batteries for the drones.

6.4 Adopted Technology and Hardware

To ensure feasibility and practicality, the proposed system relies on established technology and commercially available hardware. Table 8 outlines the adopted networking protocols.

Table 8: Networking Protocols Adopted by the System

| Protocol | Abbreviation | Description |
|---|---|---|
| Dynamic Host | DHCPv6 | Automatically provides network |

| Configuration Protocol version 6 | | configurations (IP address, subnet mask, gateway) to a host. |
|---|---|---|
| User Datagram Protocol | UDP | Transport layer protocol for efficient data transmission. |
| Internet Protocol version 6 | IPv6 | Network layer protocol enabling a large address space and improved routing. |
| Institute of Electrical and Electronics Engineers 802.11a | IEEE 802.11a | Wireless networking standard for high-speed data transmission. |
| Worldwide Interoperability for Microwave Access | WiMAX | Broadband cellular network standard for long-range wireless connectivity. |
| 802.11e Enhanced Distributed Channel Access | 802.11e EDCA | Provides Quality of Service (QoS) for prioritizing critical data traffic. |

Following established research [13-14], we adopt the DJI Phantom 4 Pro V2.0 as the UAV platform for both LD and SDs. Table 9 summarizes the drone's specifications [14].

Table 9: Specifications of the DJI Phantom 4 Pro V2.0 Drone

| Category | Specifications |
|---|---|
| Aircraft and Camera | Weight: 1375 g, Diagonal Size (Propellers Excluded): 350 mm, Max Ascent Speed (S-mode): 6 m/s, Max Ascent Speed (P-mode): 5 m/s, Max Descent Speed (S-mode): 4 m/s, Max Descent Speed (P-mode): 3 m/s, Max Speed (S-mode): 72 km/h, Max Speed (A-mode): 58 km/h, Max Speed (P-mode): 50 km/h, Max Wind Speed Resistance: 10 m/s, Max Flight Time: Approx. 30 minutes, Satellite Positioning Systems: GPS/GLONASS, Hover Accuracy Range (with GPS Positioning) Vertical: ±0.5 m, Hover Accuracy Range (with GPS Positioning) Horizontal: ±1.5 m, Camera Sensor: 1-inch CMOS, Effective pixels: 20M, Max Video Bitrate: 100Mbps, Supported SD Card: microSD, Capacity: 128GB |
| Infrared Sensing System | Obstacle Sensory Range: 0.2-7 m, FOV (Horizontal): 70°, FOV (Vertical): ±10°, Measuring Frequency: 10 Hz, Operating Environment: Surface with diffuse reflection material, and reflectivity > 8 percent (such as walls, trees, humans, etc.) |
| Intelligent Flight Battery | Capacity: 5870 mAh, Voltage: 15.2 V, Battery Type: LiPo 4S, Energy: 89.2 Wh, Net Weight: 468 g, Charging Temperature Range: 5° to 40°C, Max Charging Power: 160 W |

Table 10 details the payload elements attached to the drones, along with their weights.

Table 10: Drone's Payload Elements and Their Weights

| Payload Element | Weight (g) | LD Payload | SD Payload |
|---|---|---|---|
| Raspberry Pi 3 B | 42 | Yes | Yes |
| Arducam OV5647 Camera | 9 | No | Yes |
| Li-Polymer Battery HAT for Raspberry Pi, SW6106 Power Bank Solution | 76 | Yes | Yes |
| EHAO 104060 3000mAh Lipo Rechargeable Battery (extra battery) | 48 | Yes | Yes |
| Navio2 Autopilot Flight Controller | 23 | Yes | Yes |
| SIM7600E HAT WiMAX Adapter | 12.6 | Yes | No |
| Total Weight (g) | 201.6 | | |
| Total Weight (g) | 198 | | |

The Raspberry Pi serves as the onboard processing unit. The camera is used by SDs for data collection. A battery hat with a 3000mAh battery powers the Raspberry Pi, and an additional 3000mAh battery provides extra capacity. The Navio2 flight controller enables the Raspberry Pi to control the drone's flight, and the WiMAX adapter facilitates long-range communication between the LD and the DMC.

**7. Conclusion and Future Work**

This paper presented a novel and resilient SMI architecture leveraging a self-organizing UAV swarm for efficient, scalable, and reliable smart meter data collection. Our system integrates robust failure-handling mechanisms, dynamic task allocation strategies, and advanced communication protocols to ensure continuous operation in dynamic and unpredictable environments. Extensive simulations validated the system's effectiveness, demonstrating high formation accuracy, rapid failure recovery times, and the ability to adapt to various challenging scenarios.

Future research will focus on further enhancing system capabilities:

1. Decentralized Control Algorithms: Exploring fully decentralized control algorithms to enhance swarm autonomy, eliminate the single point of failure associated with the LD, and enable more complex, distributed decision-making within the swarm.
2. Communication Optimization for Large Swarms: Investigating and implementing adaptive communication protocols that dynamically adjust data transmission rates and frequencies based on swarm size, distance from the DMC, and environmental conditions to minimize communication overhead and extend operational range.
3. Advanced Failure Prediction Using Machine Learning: Incorporating machine learning algorithms trained on historical data and real-time sensor readings to develop more sophisticated failure prediction models for individual drones. This could enable proactive maintenance, optimize drone utilization, and further enhance system reliability.

4. Integration with Other Smart City Sensors and Systems: Exploring the integration of our proposed SMI system with other smart city sensors and systems, such as those used for traffic monitoring, environmental sensing, and infrastructure inspection, to create a more holistic and interconnected urban data platform.

By addressing these research avenues, we can unlock the full potential of UAV swarms for revolutionizing smart city applications, contributing to the development of more efficient, sustainable, and resilient urban environments.